\documentclass[12pt]{iopart}
\usepackage{graphicx}
\begin{document}

\title[Search for Inspiraling Neutron Stars in LIGO S1 data]
{Binary Neutron Star Inspiral Search in LIGO S1}

\author{Gabriela Gonz{\'a}lez, for the LIGO Science Collaboration 
\footnote{This paper was presented in the 5th Edoardo Amaldi
Conference on Gravitational Waves, by Gabriela Gonz{\'a}lez, to whom
correspondence should be addressed (gonzalez@lsu.edu)}}

\address{Department of Physics and Astronomy, 
Louisiana State University, Baton Rouge, LA, 70803}

\begin{abstract}

We describe the search for gravitational waves from inspiraling
neutron star binary systems, using data from the first Scientific Run
of the LIGO Science Collaboration. 

\end{abstract}


\section{Introduction}
In 2002, the LIGO Scientific Collaboration and the LIGO laboratory
organized the first science data run (S1). The Collaboration took data
for 17 days, between 23 August and 9 September, using the three LIGO
detectors (two detectors in the Hanford Observatory and one detector
in the Livingston Observatory), and the GEO detector in Hannover,
Germany. The detectors had not achieved their aimed sensitivities, and
were at different completion levels with respect to their
configurations, as described in \cite{LIGOS1instpaper}. However, the
noise level of at least some of the detectors was low enough to make
them sensitive to inspiraling binary neutron stars in the Galaxy, and
even the Magellanic clouds, making the data taking and analysis effort
worthwhile and competitive with previous searches that produced upper
limits on the rate of inspiraling binary neutron sources in the Galaxy
(\cite{TAMA,40m}). 

The results of the analysis of S1 LIGO data looking for gravitational
waves from binary neutron stars were described in detail in
\cite{S1Inspiral}. This article reports on some of the details of the
data analysis done in \cite{LIGOS1instpaper}, 
on the results obtained, and on some of the lessons learned
that will lead to improved methods in the analysis of present and
future science runs. At the time of writing, the Collaboration is
analyzing the data from another 2-month science run, and is preparing to
take data again with another 60+ days long run at the end of 2003.

\section{Detectors' sensitivity to binary neutron star systems in S1}

\begin{figure}
\begin{center}
\includegraphics[height=3in]{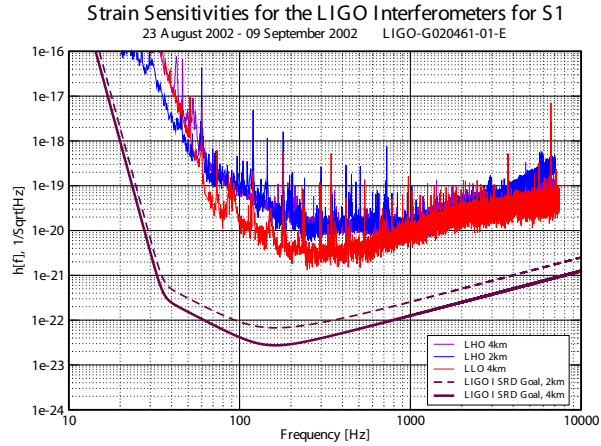}
\caption{\label{S1asd} Typical sensitivities of the LIGO detectors during S1.}
\end{center}
\end{figure}

A typical amplitude spectral density of the LIGO detectors' noise
during S1, interpreted as strain, is shown in Fig.~\ref{S1asd}. We can
translate the noise spectral density into an optimal range, defined as
the maximum distance at which a binary neutron star system, if
optimally oriented and located, would be detected with a
signal-to-noise ratio of 8. In average, the optimal range for the
Livingston detector during S1 was 176 kpc, and for the Hanford 4km
detector, 46 kpc. The optimal range for the 2km Hanford detector was
slightly worse than for the 4km detector, and the data quality was not
as good; the range for the GEO detector for the same sources was
significantly worse. We then decided to use only data from the LIGO
4km detectors, which we call L1 (Livingston detector) and H1 (Hanford
detector).

Seismic noise at the Livingston Observatory limited operations during
most weekdays, so the overall L1 duty cycle during S1 was 42\% (170
hours of data), while the duty cycle of H1 was 58\% (235 hours). The
time when the two detectors H1 and L1 were in operation amounted to
116 hours, representing a duty cycle of only 28\%. We decided to
analyze {\it all} of the data available from L1 and H1, including the
times when a single detector was available. While we were not going to
be able to confirm through coincidence the candidate events found when
only one detector was operating, in this way we have more statistics
to use for upper limit analysis and for testing our data analysis
method. In the future, with longer science runs and improved duty
cycles, we will be able to achieve good upper limits on astrophysical
rates using only times when more than one detector is in operation.

The scientific goal of the data analysis exercise was to set an upper
limit on the rate of binary neutron star systems in the Galaxy. We did
not observe any coincident event with signal-to-noise ratio (SNR)
larger than 6.5 in both detectors; thus, we had no candidates for
detection. 

In order to get an upper limit on the event rate, we measure the
efficiency of the detectors (and the search method) to the population
we are sensitive to (Milky Way Galaxy, Large and Small Magellanic
Clouds), and infer the upper limit rate in that population from the
efficiency as a function of signal-to-noise $\epsilon(\rho)$ and the
amount of time analyzed $T$. The result is an observational,
frequentist, upper limit of 170 events/year per Milky Way Equivalent
Galaxy (MWEG), with 90\% confidence, on the coalescence rate of binary
systems in which each component has a mass in the range 1-3 M$_\odot$.

A rough estimate of what could be expected from the analysis shows
that our result was close to expectations. At SNR=16 (the SNR of our
strongest surviving candidate), L1 was, in average, sensitive to 90\%
of the sources in the Galaxy, while H1 was sensitive to 40\% of the
same population. L1 was operating for 170 hours in S1, while H1
provided an additional 119 hours operating without L1 in
coincidence. The total of available hours was then 289 hours. The
maximum efficiency expected was then $(0.9\times 170+0.4\times
119)/289=0.69$. An upper limit on the event rate for the Galaxy is
then expected at the level of $2.3/\epsilon T \sim$ 100
events/yr. This is considerably smaller than the actual upper limit
obtained of 170 events/yr. This was due to several reasons: not all
the data was used because of availability, calibration problems or
poor data quality; there was some loss of efficiency due to
requirements on matching templates and on consistency under
coincidence; and we used the upper bound in our rate estimate given
the uncertainties in the results. Of course, even an optimally
calculated event rate would be far from other astrophysical estimates
which suggest rates of $10^{-5}$/yr for our Galaxy\cite{Rates,Kim:2002uw}.

\section{Data analysis method and results}

We developed a pipeline to analyze the data that could lead to
detection during times when the two detectors were in operation,
but also analyzed the data when a single detector was in operation.
If this pipeline had resulted in a strong candidate appearing in both
detectors, consistent with a coincident signal, we would have 
followed up with additional investigations to determine its origin,
either astrophysical or instrumental. In S1, our analysis did not find
any such coincident event. In the absence of detection, we take
several more steps to produce, from the list of candidates, an
observational upper limit on the rate of events in the Galaxy. We
describe now the different elements leading to our resulting rate of
170 events/yr/MWEG.

The pipeline and details on the methods used for each step are
described in detail in \cite{S1Inspiral}; here we summarize the steps
and point out possible improvements to apply in the analysis of future
data.

\subsection{Candidate events}

The pipeline has several parameters that were tuned to obtain the best
upper limit possible. Since biases could be introduced by this tuning,
we performed the optimization on a selected data set, called the
``playground''. This playground, about 10\% of the data in
coincidence, was not used in the final analysis. The playground times
were chosen by hand, trying to choose locked segments that were
representative of the different data quality over the run. However,
a fair representation was difficult to obtain with such an ad-hoc
procedure, and for the next science run (S2, Jan 14-Apr 14 2003), an
automatic procedure was implemented.

In the first step in our pipeline, we analyze the interferometer data
from each detector using matched filtering, with a template bank
chosen to guarantee coverage of the worst detector \cite{S1Inspiral}
(and thus producing overcoverage of the better detector). The matched
filtering makes critical use of the detector data calibration: we lost
about 25 hours, or 9\% of available data, due to missing calibration
information. We also lost 39 hours, or 13\%, due to the granularity of
the matched filtering jobs, which required blocks of time 256 seconds
long, and thus could not use short segments, or times at the end of
segments. We expect to improve both of these numbers with better
practice (for calibration), and better instrument stability (for
longer operating segments).

We used a single template bank for both detectors, for all times in
S1. In future runs, we will use different banks, dynamically adapted
to the noise in the detector in the segment of continuous lock being
analyzed. When the SNR of a particular candidate is larger than 6.5,
we calculate a quality-of-fit parameter (similar to a $\chi^2$, but
adapted to the limitations placed by a finite template bank), and tune
a cut on this parameter. This cut, although very powerful, was found
not to veto candidates that upon further inspection were happening
during noisy times. We are now developing methods that will
analyze the time before the candidate template starts, either looking
for excitations of other templates (typical during noisy times), or
looking for lack of consistency with a stationary noise background.

The second step in the pipeline applied two kind of instrumental
vetoes to the surviving triggers: an epoch cut, and an instrumental
veto for H1. The epoch cut was adopted by the two working groups
looking for signals of short duration (bursts and inspiral sources),
and eliminated times when the noise in certain frequency bands was
well outside typical noise levels during S1, trended over 6 minute
periods. This cut removed 8\% of the L1 data, and 18\% of the H1
data. In spite of the non-negligible amount of data eliminated, we did
not have a satisfactory explanation on the reasons why the noise was
excessively high at these times. This was in part due to the lack of
time to diagnose the noise sources: the instruments are constantly
changing, except during the data taking run itself. This makes the
diagnostics of particular data features very difficult, especially if
attempted long after the instrument has changed character. We hope the
understanding of the instrument will be better when we reach a
stationary (and satisfactory) noise level, but probably not earlier.
The second instrumental veto was due to coupling of frequency noise
glitches in H1, as seen in another interferometer channel
simultaneously with triggers in the gravitational wave channel.
We had similar efficient veto channels for L1 (which also
had a much higher event rate of accidental triggers), but we decided
not to use them because when injecting hardware signals to simulate
gravitational waves, they also excited the auxiliary channels in ways
that could produce ``vetoes''. We expect that with better understanding
of the physical nature of the vetoes, we will be able to find
efficient, safe vetoes that will eliminate the candidate events
produced by instrumental artifacts. 

The third step involves only the candidate triggers in L1 surviving
the template matching and instrumental vetoes applied, and which are
strong enough to show up in H1 with SNR$>$6.5. To these events, we
apply a coincidence veto if the are no consistent triggers in H1 and
L1. However, there were no such events in the S1 data: only triggers
in L1 which appeared closer than 51 kpc would have possibly appeared as
candidates in H1; but the strongest candidate event we had during
coincident times had an apparent distance of 68 kpc. During S1, we
were logistically limited in the number of triggers generated, thus
imposing a lower limit of SNR looked for as 6.5.
Thus, the lower limit in SNR for the noisier detector sets up a much
stricter SNR criterion in the less noisy detector.
To overcome this problem without
overloading the database with low SNR events, we plan to implement a
hierarchical search, looking for low SNR in the less sensitive
detector only around times when a large SNR event is found in the more
sensitive detector. 

The final step in the pipeline generating the list of candidates
involved maximizing all surviving triggers over time and over the
template bank, to improve the timing resolution, and the confidence in
the SNR of the candidate. This pipeline analyzed a total of
236 hours.

\subsection{Detector efficiency}

To measure the efficiency of our pipeline, we injected in software
simulated signals, following the sample population for spatial and
mass distributions from a Milky Way population produced by the
simulations of Ref.~\cite{Belczynski:2002}, with the spatial
distribution described in Ref.~\cite{Kim:2002uw}.  Since L1 was
sensitive to sources at a distance slightly larger than our Galaxy, we
added sources from the Large and Small Magellanic Clouds, treating
them as points at their known distances and sky positions. These
systems contributed about 11\% and 2\% of the rate of the simulated
signals, respectively. The total number of injected signals totaled
more than 5000.
The efficiency is also measured as a function of SNR: at each SNR, we
measure the number of simulated sources in our model at the
corresponding distance or closer. We measured an efficiency of 80\%
for SNR$>$8, and 53\% for SNR=15.9, that of our loudest surviving
candidate.

\subsection{Upper Limit Result}

In principle, we would like our pipeline to produce some non-zero rate
of candidate events as a function of threshold SNR, and compare that with an
estimated accidental rate at the same SNR. The background rate can be
measured by time-shift analysis, which guarantees that any coincidence
is accidental. However, our pipeline did not result in {\em any} event
for which coincidence timing criteria could be applied, making this
method for estimating the background impossible. For single detector
data, no such reliable estimate of the background can be performed: we
just have a list of candidates which we cannot confirm or veto through
coincidence.  Assuming the true sources have a Poisson distribution
with rate $R$, then the probability $P$ of observing a signal with
SNR=$\rho$ in a time of observation $T$, with the efficiency of the detector and method as a function of SNR given by $\epsilon(\rho)$, is
$P=1-e^{-RT\epsilon(\rho)}$. Given a list of candidate events with
maximum SNR=$\rho_*$, we interpret that with confidence level $P$, the
rate is not larger than $-ln(1-P)/\epsilon(\rho_*)T$, or
$2.3/\epsilon(\rho_*)T$ for 90\% confidence. 

Given our largest SNR=15.9 surviving in the L1-only pipeline, our
measured efficiency $\epsilon(\rho=15.9)=0.53$ to the simulated
population, and our runtime $T$=236 hr, we obtain a rate $R\leq 161$
events/year. If we express the efficiency to sources in our Galaxy,
comprising 88.5\% of the population considered, we obtain an upper
limit on the rate of 140 events/year/Milky Way Equivalent Galaxy. We
have +14\%/-10\% uncertainties in the efficiency, mostly due to
calibration uncertainties, and 5\% uncertainties in the population
estimates; we choose to state the obtained upper limit as the higher
bound in our rate estimate, or 170 events/year/MWEG.

\section{Conclusions}

For the first time, we used data from coincident interferometric
detectors to look for signals from binary inspiraling neutron star
systems. We did not find any signal with SNR$>$6.5 while the two LIGO
detectors were in operation. We used the data to set upper limits on
the possible rates of inspirals in our Galaxy. These are very
important successful milestones in the first steps towards the data
analysis of interferometric data, and we have learned many important
lessons from the exercise which we plan to use in the future analysis
of science runs.


\section*{Acknowledgments}
We gratefully acknowledge the LIGO project and the LIGO Science
Collaboration, who made the first LIGO Scientific Run possible. This
work was supported by NSF grant PHY-0135389.

\end{document}